\documentclass{ws-ijmpe}

\begin{document} 

\markboth{Rudolph C. Hwa}{Hadron Correlation in Jets}

\title{HADRON CORRELATION IN JETS}

\author
 {RUDOLPH C. HWA}
 
\address
{Institute of Theoretical Science and Department of
Physics\\ University of Oregon, Eugene, OR 97403-5203, USA}

\maketitle

\begin{abstract}
We review some recent experimental and theoretical work on the correlation among hadrons produced at intermediate $p_T$ at RHIC.  The topics include:  forward and backward asymmetry with and without trigger at mid-rapidity, associated-particle distribution on the near side, the $\Omega$ puzzle and its solution, associated particles on the away side, and two-jet recombination at LHC.
\end{abstract}

\section{Introduction}

In heavy-ion collisions the properties of jet structure are very different from those of jets produced in $e^+e^-$ and hadronic collisions.  The effects of the dense medium on hard scattering of partons drastically alter the features of parton fragmentation in vacuum for $p_T < 8$ GeV/c.  Nowhere is it more evident than in the $p/\pi$ ratio that exceeds $1$ at $p_T \sim 3$ GeV/c in central Au+Au collisions,\cite{sa,ba} a phenomenological fact that cannot be reproduced by parton fragmentation, even if medium-modified.  Most of the data on jet correlation are in the same $p_T < 8$ GeV/c region.  Quantitative determination of hadronic correlations in jets has mainly been done in the framework of recombination.  We give here a brief review of the latest development in that subject.

An earlier review of correlation in jets was given in April 2005 at the MIT Workshop on Correlations and Fluctuations.\cite{CF,rch}  We shall not repeat what is contained there.  One other topic that we shall not include here is autocorrelation.\cite{hy}  Most correlation studies involve a trigger and one or two associated particles.  Autocorrelation involves two particles treated on the same footing, and has no need for background subtraction.\cite{tat,tat2,ja}  However, for lack of space and high-$p_T$ data, this interesting topic will be left out.

\section{Forward-backward  asymmetry in d+Au collisions}

We start with the asymmetry  in high-$p_T$ yields in the forward vs backward production in d+Au collisions, first on uncorrelated multiplicities and then on multiplicities correlated to trigger at mid-rapidity.  For a long time the Cronin effect\cite{jc} has been interpreted as being due to the initial-state transverse broadening of partons in the deuteron as they traverse the target nucleus, but we have shown that the final-state interaction at hadronization gives rise to enhancement at high $p_T$ for both pion and proton production.\cite{hy2}  Now, for forward-backward asymmetry, if the initial-state broadening is important, then forward ($d$ side) production has more transverse broadening than backward (Au side), since $d$ is the small nucleus that does not give much broadening to any parton in Au  in the backward direction.  Thus one would expect more forward $(F)$ particle produced at high $p_T$ than backward $(B)$ particles, i.e., $B/F < 1$.  But the data show the opposite,\cite{ja2} i.e., $B/F >1$ for all $p_T$ examined, up to 6 GeV.  See Fig.\ 1(a).  The range of $\eta$ is $0.5 < |\eta| <1.0$.

\begin{figure}[th]
\hspace*{-.1cm}{\psfig{file=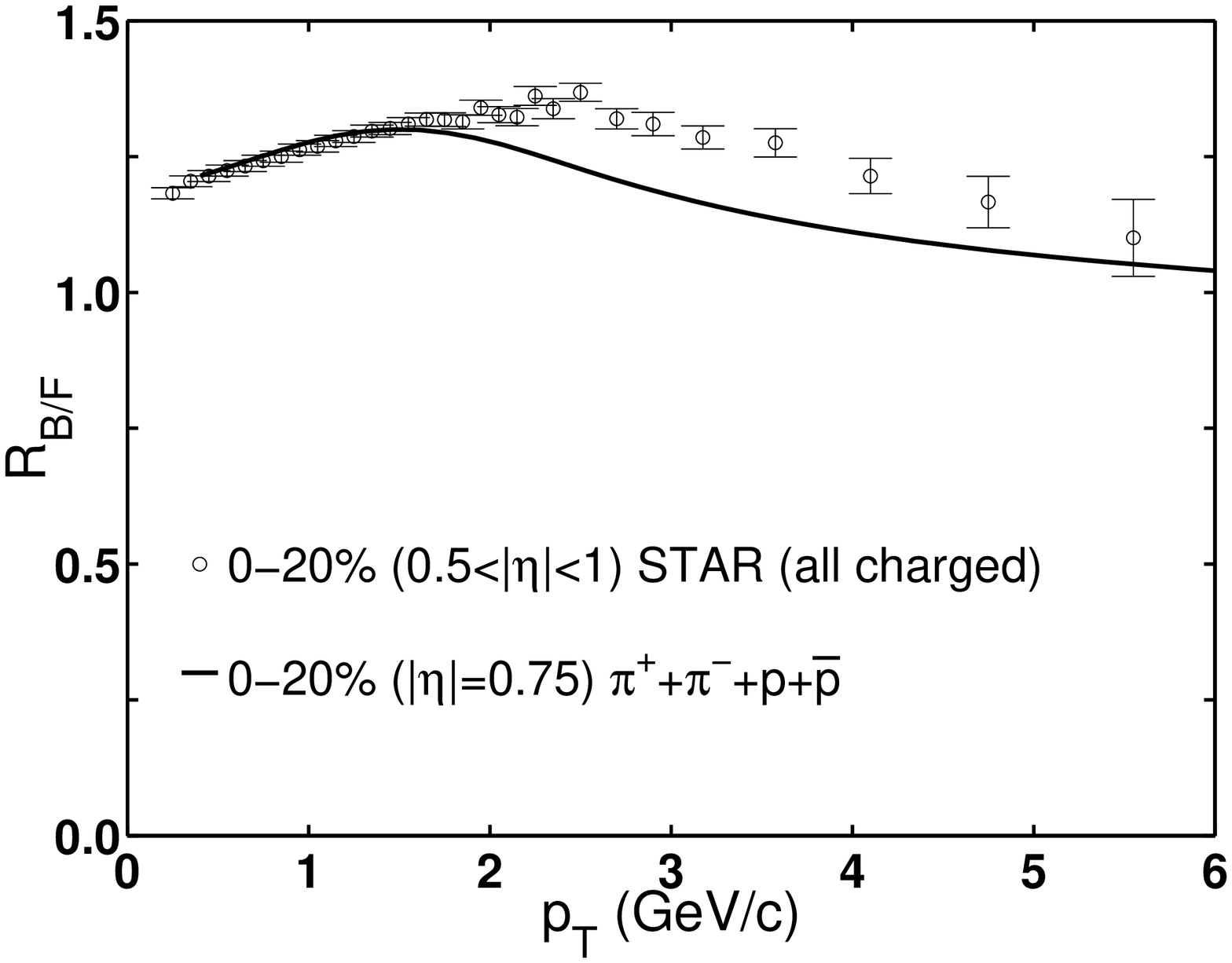,width=6cm}}
\end{figure}

\vspace*{-5.9cm}
\begin{figure}[th]
\hspace*{6cm}{\psfig{file=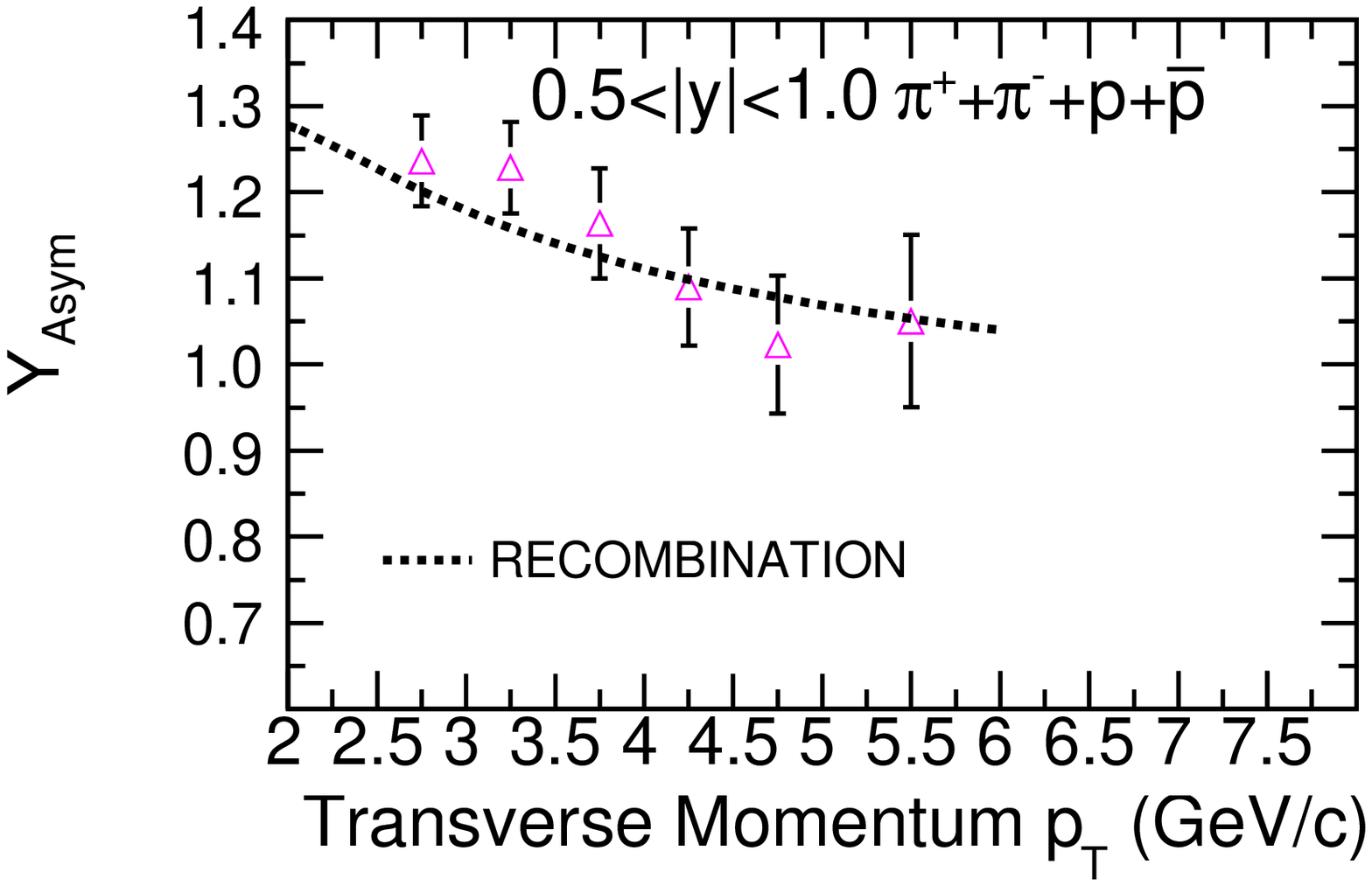,width=7.4cm}}
\end{figure}
\begin{figure}[th]
\vspace*{-1cm}
\caption{(a) Ratio $B/F$ calculated in the RM [12] (solid line), compared to data [11]. 
\hspace*{2.15cm}(b) The same $B/F$ ratio with data from Ref.\ [13].}
\end{figure}

In Ref.\ [12] we studied the problem in the recombination model (RM).  The important input is the rapidity distribution that shows enhancement in the $B$ direction relative to the $F$ direction for obvious reason.  That is shown below in  Fig.\ 2(b).  Since soft partons have $B/F > 1$, the recombination of thermal and shower partons naturally leads to $B/F > 1$ also at moderate $p_T$.  The solid line in Fig.\ 1(a) is the result of our calculation, which agrees with the data well up to $p_T \approx 2$ GeV/c, but is below the data for 
$p_T > 2$ GeV/c.  However, subsequent reanalysis of the data shows that the experimental points have come down slightly  to agree with our prediction very well, as shown in  Fig.\ 1(b).\cite{ba2}

We now go to the more recent data on the correlation between mid-rapidity trigger and associated particles at forward or backward rapidities in d+Au collisions.\cite{fw}  In our view the dynamical origin of the asymmetry is the same as above, independent of the trigger at mid-rapidity.  In  Fig.\ 2(a) is shown the STAR data on the $\Delta \phi$ distribution for $B$ in red $(- 3.9 < \eta < - 2.7)$ and $F$ in blue $(2.7 < \eta < 3.9)$.  The peaks are centered at $\Delta \phi = \pi$, and the yields are roughly $B/F \approx 2$.  This ratio is much larger than that shown in Fig.\ 1.  That is understandable, since in Fig.\ 1 the range of $\eta$ is $0.5 < |\eta| <1.0$, whereas in Fig.\ 2(a) it is $2.7 < |\eta| <3.9$.  For such a large difference in the rapidity ranges, there is a correspondingly large difference in the density of soft partons, as can be seen in the rapidity distribution of soft hadrons in  Fig.\ 2(b).  The thermal-shower recombination that is responsible for what is seen in  Figs.\ 1(a) and 2(a) depends linearly on the thermal-parton density, whereas the thermal-thermal recombination that is dominant for the distribution in  Fig.\ 2(b) depends quadratically on the thermal-parton density.  In the former case if we take the ratio of $B/F (2.7<|\eta|<3.9)$ to $B/F ( |\eta|<0.75)$, i.e. 2.2 at the peaks in Fig.\ 2(a) to 1.3 at the peak in Fig.\ 1(a), we get $R_{TS} = 1.7$.  In the latter case if we take the ratio of $B/F (|\eta|=3.2) = 3.4$ to $B/F (|\eta|=0.75) = 1.16$, both for 0-20\% in Fig.\ 2(b), we get $R_{TT} = 2.9$.  We see that $R_{TT}$ is indeed nearly the square of $R_{TS}$.

\vspace*{0cm}
\begin{figure}[th]
\hspace*{0cm}{\psfig{file=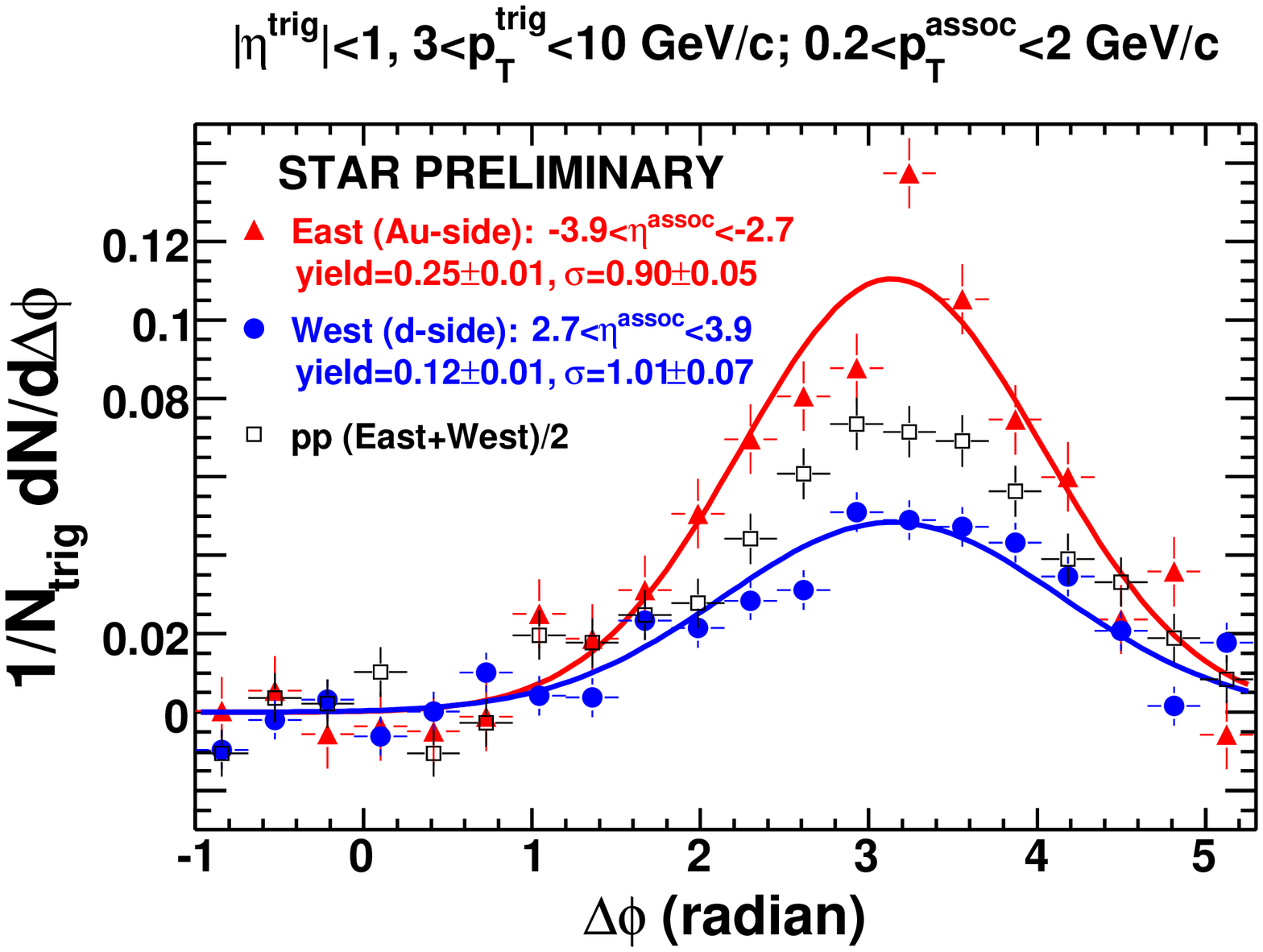,width=6.5cm}}
\end{figure}
\vspace*{-6.5cm}
\begin{figure}[th]
\hspace*{7cm}{\psfig{file=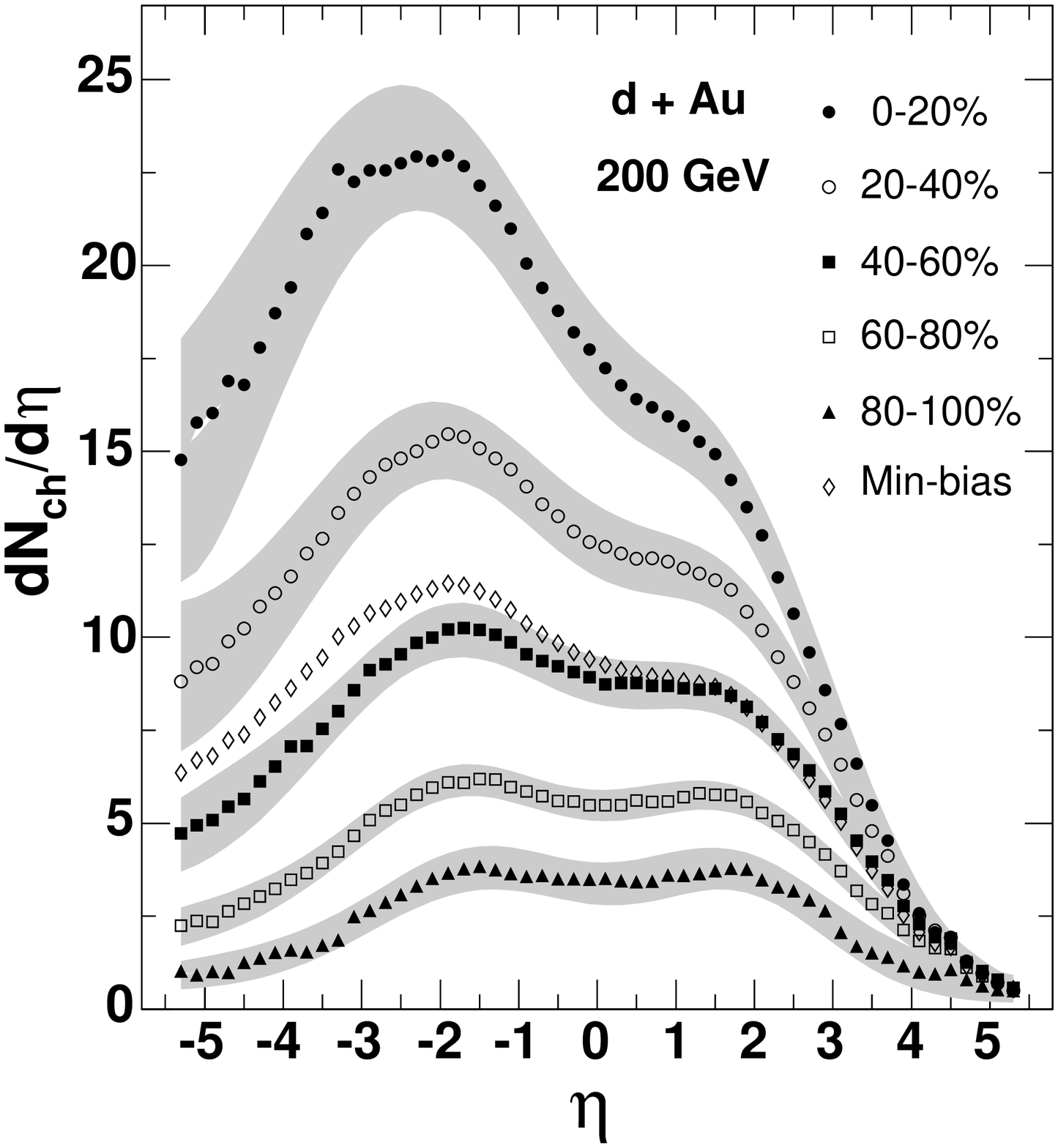,width=5.5cm}}
\end{figure}
\begin{figure}[th]
\vspace*{-1cm}
\caption{(a) $\Delta\phi$ distribution for $B$ (red) and $F$ (blue) associated with trigger at mid-$\eta$ [14]. \hspace*{1.1cm}(b) Pseudo-rapidity distribution of charged particles in d+Au collisions [15].}
\end{figure}
\vspace*{-.5cm}

Thus we conclude that in d+Au collisions the correlation of forward or backward particles associated with a trigger at mid-rapidity is understandable in the framework of the RM, and is consistent with the earlier observation of forward/backward asymmetry without the necessity to invoke saturation physics.\cite{hyf} 

\section{Associated-particle $p_T$ distribution on the near side}

In the RM the $p_T$ distributions of pions associated with pion triggers at various centralities in d+Au and Au+Au collisions have been calculated in.\cite{ht}  It was found that there is very little dependence on centrality in d+Au collisions, but significant dependence in Au+Au collisions.  Fig.\ 3(a) shows the central-to-peripheral ratios in red and blue  lines for d+Au and Au+Au collisions.  They agree approximately with the data  shown  in Fig.\ 3(a).  For the ratio $AA/pp$ there is an enhancement factor of about 3 at $p_T \approx 1$ GeV/c,\cite{ja3} which, in RM, is the result of thermal-shower recombination.

More recently, higher statistics data of STAR were analyzed to give $0$-$10\%/40$-$80\%$ ratio for $h$-$h$ correlation with very small error bars for $3 < p^{\rm{trig}}_T < 6$ GeV/c.\cite{jb}  The new result is in fair agreement with the blue line of Fig.\ 3(a) calculated earlier, as shown in Fig.\ 3(b), considering the fact that the peripheral bins are not exactly the same.

\begin{figure}[th]
\hspace*{-0cm}{\psfig{file=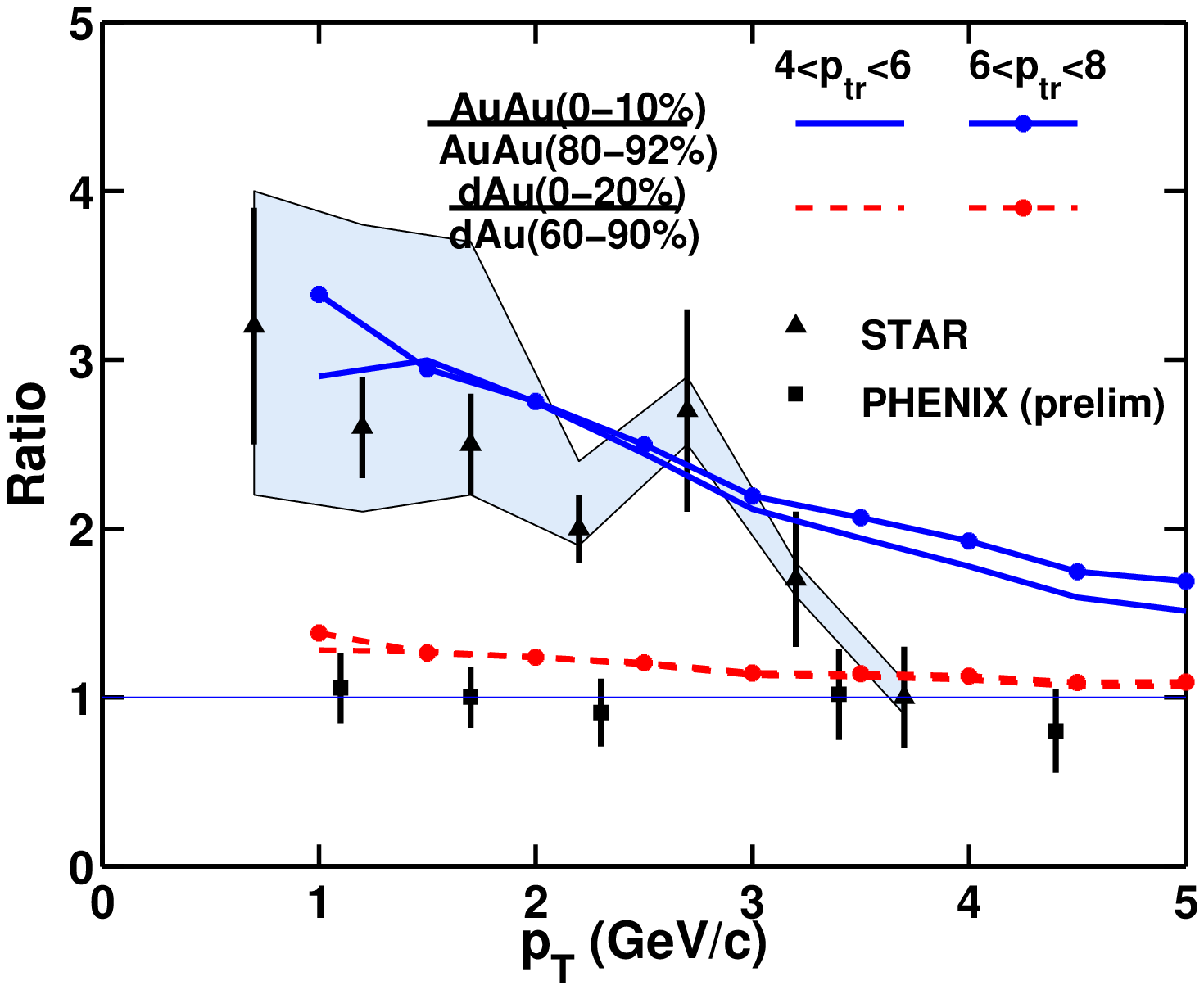,width=6.5cm}}
\end{figure}
\vspace*{-6.3cm}
\begin{figure}[th]
\hspace*{6.3cm}{\psfig{file=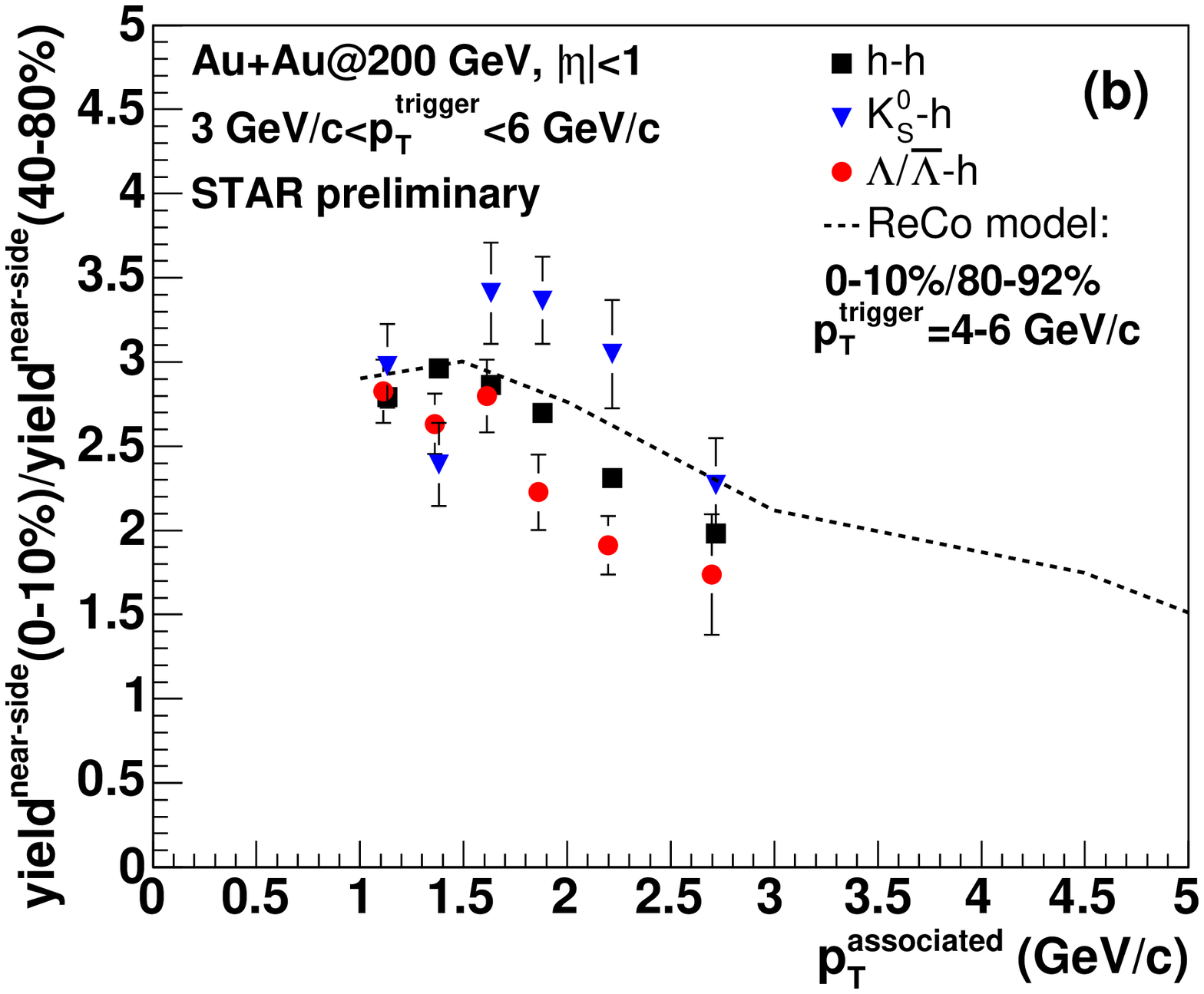,width=6cm}}
\end{figure}
\begin{figure}[th]
\vspace*{-1cm}
\caption{(a) $R_{CP}$ of associated particle distributions. Lines are calculated in the RM [16]. \hspace*{1.6cm}(b) More recent data from STAR on the ratio [18].}
\end{figure}
\vspace*{-.5cm}
Ref. [18] also has data on the correlation ratios for $K_s$ and $\Lambda$ triggers.  No calculations have been done yet for those correlations involving strange particles.  It does not seem that there is any difficulty in principle, although in detail there are some subtleties with strange triggers, the most puzzling one of which is the $\Omega$ particle.  We go to the $\Omega$ puzzle next, after some overview of jet phenomenology.

\section{The case of the vanishing jet and the $\Omega$ puzzle}

The jet structure on the near side is now well known to have a peak and a ridge.  They have been referred to as Jet (J) and Ridge (R).  Putschke has shown that for fixed trigger momentum both J and R yields decrease exponentially with $p^{\rm{assoc}}_T$; however, the J yield increases with $p^{\rm{trig}}_T$, while the R yield is relatively independent of $p^{\rm{trig}}_T$.\cite{jp}  For $p^{\rm{trig}}_T > 6$ GeV/c the ratio J/R is roughly $>1$, especially at lower  $p^{\rm{assoc}}_T$.  For $h$-$h$ correlation it was shown by Bielcikova that J/R is about 10-15\% when $3 < p^{\rm{trig}}_T < 6$ GeV/c and $p^{\rm{assoc}}_T$ is as low as 1 GeV/c.\cite{jb2}  For $\Lambda$-$h$ correlation J/R seems to be another factor of 3 lower, i.e. $\sim 0.04$.  Presumably, for $\Xi$-$h$ and $\Omega$-$h$ correlations J/R would be even lower.  It then becomes a rather intriguing situation where J disappears and the only visible structure that remains of a jet is the Ridge.  Let us call it a phantom jet, a notion that we shall refer to below when we discuss the $\Omega$ puzzle.

The production of strange particles at intermediate $p_T$ has been studied in the RM in Ref. [21].  For particle like $K$ and $\Lambda$ that contain $u$ and $d$ quarks, it is the shower light quarks that boost the $p_T$ of those hadrons.  The shower $s$ quarks are suppressed at intermediate $p_T$, so it is the thermal $s$ quarks that add the strangeness flavor to them.  That kind of TS recombination is not important for $\phi$ and $\Omega$ that contain no light quarks.  Only the recombination of thermal $s$ quarks can contribute to the dominant component of $\phi$ and $\Omega$.  That means that the $p_T$ distributions of $\phi$ and $\Omega$ should be exponential, which is the characteristic of thermal partons; indeed, that is what has been observed.\cite{ja4,sb}   In Fig.\ 4(a) is shown the $p_T$ distribution of $\Omega$ that is well reproduced by the recombination of thermal partons only, the TTT term being completely covered up by the solid line that is the sum of all terms.

Since thermal partons are mainly statistical, it is natural to conclude that any particle associated with a $\phi$ or $\Omega$ trigger is in the statistical background.  That conclusion, stated in,\cite{hy3} led the STAR collaboration to an intensive study of the $\Delta \phi$ distribution on the near side of $\Omega$ trigger.   The result was reported at Quark Matter 2006.\cite{jb3} Fig.\ 4(b) shows that there are particles associated with $\Omega$ above the $v_2$ background.  At face value that figure contradicts the prediction in Ref.\ [21].

\begin{figure}[th]
\hspace*{-0cm}{\psfig{file=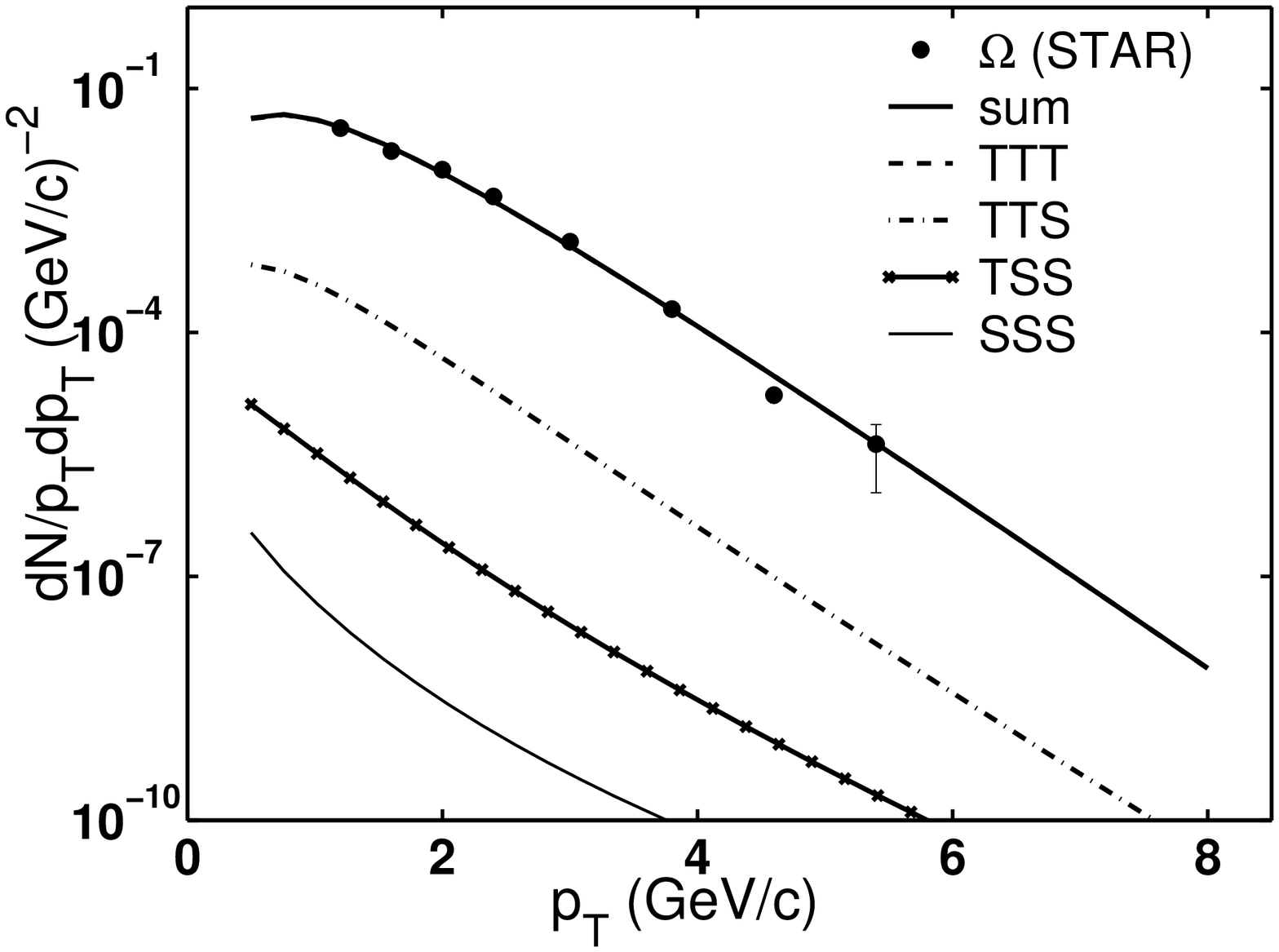,width=5.5cm}}
\end{figure}
\vspace*{-5.4cm}
\begin{figure}[th]
\hspace*{5.9cm}{\psfig{file=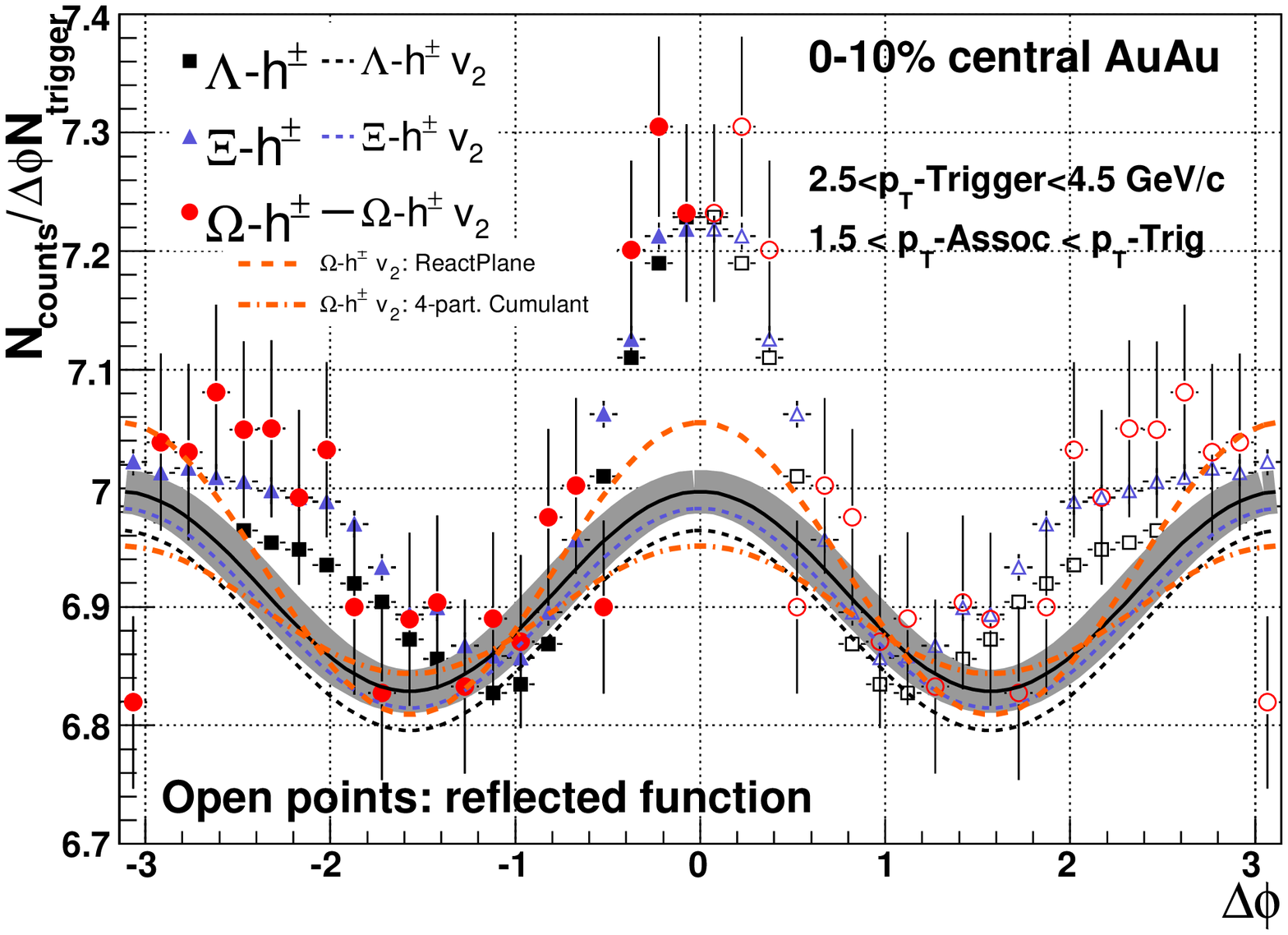,width=6.7cm}}
\end{figure}
\begin{figure}[th]
\vspace*{-1cm}
\caption{(a) $\Omega$ spectrum in central  Au+Au collisions. The lines are from [21], the data from [22,23]. (b) Recent data from STAR [24] on the $\Delta\phi$ distributions of charged particles associated with $\Omega$, and also $\Lambda$ and $\Xi$.}
\end{figure}
\vspace*{-.5cm}

With that new result we now have a puzzle:  how can the $\Omega$ have both an exponential spectrum (i.e., from a thermal source) and also associate particles (implying jet structure)?  The solution is to be found in the recognition that it is the manifestation of phantom jet.  When the trigger momentum is not large, and the associated-particle momentum is low, it is possible for the ridge to dominate the jet structure, as discussed above.  Both the $\Omega$ and its associated particles can originate from the ridge, which is known to arise from the enhanced thermal partons due to the energy loss of a hard parton traversing the medium.\cite{ch}  Thus the $\Omega$ spectrum is exponential due to thermal recombination and is accompanied by associate particles also from the ridge that is above the statistical background.\cite{rch2}

To verify this solution of the puzzle there are three tests that it should pass.  First, the $\Delta \eta$ distribution should be determined at the same $p^{\rm{trig}}_T$ and $p^{\rm{assoc}}_T$ ranges as in $\Delta \phi$, and be found to consist of only the ridge mainly, thereby proving that the excess over background in the $\Delta \phi$ distribution arises only from the ridge.  Second, the associated-particle distribution in $p_T$ should be exponential.  Third, the $p/\pi$ ratio among the associated particles should be large, of order 1 rather than 0.1.  These are all characteristics of thermal-parton recombination.\cite{rch2}

Along the same line of reasoning it may be possible to understand the puzzling difference between the baryon-triggered and meson-triggered yields of meson partners in central collisions.\cite{as}  For $2.5 < p^{\rm{trig}}_T < 4$ GeV/c and $1.7 < p^{\rm{assoc}}_T < 2.5$ GeV/c it is found that meson-triggered events have nearly five times more meson partners than in baryon-triggered events at $N_{\rm{part}} = 350$, but they are nearly equal for $N_{\rm{part}} < 250$.  A possible explanation is that at higher $N_{\rm{part}}$ a larger ridge is developed and that baryon triggers at low $p^{\rm{trig}}_T$ originate mainly from the ridge due to the recombination of thermal partons only.  Meson triggers on the other hand may arise more from TS recombination since more parton momentum is needed for the same hadron momentum as a baryon.  This difference in the partonic origin of the triggers may be the source of the difference in the yields of the partners, since Jets are partners at all $p_T$, while the ones in ridges are exponentially damped for increasing $p_T$.  Whether this conjecture can quantitatively reproduce the data awaits detailed calculation.

\section{Associated particles on the away side}

Considerable attention has been given to the possibility of collective response of the medium to the passage of a parton in recoil against a hard parton that generates a jet.  Mach cone formation is one example that has stimulated extensive experimental searches discussed at this workshop.  We present here an alternative possibility that has sometimes been referred to as deflected jet.  We call it Markovian parton scattering (MPS).\cite{ch2}  Since the dip-bump structure on the away side is for $2.5 < p^{\rm{trig}}_T < 4$ GeV/c,\cite{sa2} the scattering process of the recoil parton in the medium is non-perturbative and can have trajectories that bend.   The only sensible way we know to treat such trajectories is to divide them into linear segments.  We assume that at each step the scattering angle $\alpha$ retains no memory of the past, i.e., Markovian.

In our MC simulation of MPS there are many parameters adjusted to reproduce the data, the details of which are given in Ref.\ [28]. The dynamical content consists of energy loss and step size that depend on energy and density, and random scattering in the forward cone at each step, the cone width being dependent also on energy and density.  The trajectories all start near the surface.  Those that exit are characterized by persistent bending in the same side-way direction, and are therefore relatively short.  Those that are absorbed zig-zag in the medium until the parton energy falls below 0.3 GeV.  The energy lost is thermalized and converted to pedestal distribution in $\Delta \phi$.  The exit tracks hadronize by recombination and are added above the pedestal, reproducing the dip-bump structure observed.\cite{ch2}

The principal difference between MPS and Mach cone processes is that they have one and two peaks, respectively, per trigger.  As the trigger momentum is increased, the trajectories in MPS become more straight and the two peaks in the $\Delta \phi$ distribution move closer to each other until they become a single peak at high $p^{\rm{trig}}_T$ and high $p^{\rm{assoc}}_T$.  There is no change in physics from low to high $p^{\rm{trig}}_T$.\cite{ch2}

The process that we describe here need not apply only at mid-rapidity.  Recent STAR data presented by F.\ Wang \cite{fw} show that with $|\eta ^{\rm{trig}}| < 1$ and $2.7 < |\eta^{\rm{assoc}}| < 3.9$ the dip-bump structure appears also in $\Delta \phi$ distribution for $3 < p ^{\rm{trig}}_T < 10$ and $0.2 < p^{\rm{assoc}}_T < 2$ GeV/c in Au+Au collisions.  The width of the double bumps becomes broader in more central collisions.  It is not clear how the phenomenon can be explained as a Mach cone effect, since the recoil parton moves almost as fast as the expanding front of the medium so that in the longitudinal co-moving frame the local density is not high.  In the deflected jet scenario the basic dynamical process remains the same as before, and the centrality dependence of the $\Delta \phi$ width can be understood simply as a consequence of the change in transverse size of the medium, i.e., the recoil parton has more path length to bend when the transverse area is larger.

\section{Two-jet recombination at LHC}

In our first paper on shower-shower recombination \cite{hy4} we have considered two cases, labeled 1-jet and 2-jet.  If two shower partons arise from the same jet, the recombination process is equivalent to fragmentation.\cite{hy5}   But if the two-shower partons are from two neighboring jets, then that would be a very different and unique signature of recombination.  At RHIC such SS recombination process is extremely rare; the corresponding curves shown in Ref.\ [30] were calculated for unrealistic overlap probability mainly for the purpose of illustrating the $p_T$ dependence.  However, at LHC the density of hard partons is so high that 2-jet recombination becomes a real possibility.

In Ref.\ [32] we have calculated the pion and proton $p_T$ distributions arising from the recombination of shower partons in nearby jets at LHC.  We restricted our attention to the region $10 < p_T < 20$ GeV/c, which is high enough to justify the neglect of thermal partons but low enough to have significant probability of overlap of neighboring jet cones.  A wide range of that probability $\Gamma$ has been considered.  The most striking result is that proton production is remarkably enhanced.  That is because 3 shower partons from two jets can each have roughly 1/3 the proton momentum without requiring as much momenta of the hard partons as would be the case if only one hard parton is involved.  For pion production the shower-parton momenta are higher and hence less abundant.  The $p/\pi$ ratio is estimated to vary from 20 down to 5 in the $p_T$ range considered.\cite{hy6}  That will be a very distinctive signature of the new process of hadronization.

The particles produced by the process described above will have associated partners, since jets are involved.  But at LHC many other jets are also produced in each event, so those particles associated with the trigger from 2-jet recombination are part of the background of an ocean of hadrons from other jets.  Thus there should be no observable jet structure distinguishable from the statistical background.\cite{hy6}
If this prediction is verified, one has to go to $p^{\rm{trig}}_T > p^{\rm{assoc}}_T \gg 20$ GeV/c to do jet tomography.  The study of correlation in jets will be challenging.

\section{Conclusion}

Many correlation phenomena related to particles associated with jets at moderate $p_T$ can be understood in terms of recombination.  To our knowledge there has been no calculations done on such correlations based on fragmentation. Jet quenching is supposed to be a method to learn about the properties of the dense medium.  Apart from the nuclear modification factor that is determined from single-particle spectra, not much has been learned about the medium by jet tomography.  Correlation phenomena at intermediate $p_T$ are dominated by the properties of hadronization at the final stage.  For example, we have learned nothing about the nature of phase transition of the bulk medium.  Jet tomography is a tool too restricted in spatial extent to obtain any information about the spatial variation and fluctuation  of the dynamical properties of the medium.

At LHC so many jets are produced that the notion of correlation in jets may well undergo some major changes from what we have learned at RHIC.  As collision energy has increased from SPS to RHIC and soon to LHC, the definition of a jet has changed from $>2$ to $>8$ and next to $>20$ GeV/c.  It is possible that pQCD may become better justified for certain processes, but the complexity of the non-perturbative processes is envisioned to become so overwhelming that obtaining clean results from simple probes is not at all obviously feasible at this vantage point.

\section{Acknowledgment}

The work reported here has been done in collaboration with C.\ B.\ Chiu, Z.\ Tan and C.\ B.\ Yang, whose participation has been crucial in the success of this program.  This work was supported,  in
part,  by the U.\ S.\ Department of Energy under Grant No. DE-FG02-96ER40972.


\end{document}